\documentclass[12pt]{article}
 \pdfoutput=1
\usepackage{amsfonts,amsthm,amsmath,amssymb,slashed}
\usepackage[textwidth = 500 pt, textheight = 630 pt]{geometry}
\usepackage{latexsym,amsfonts,amsmath,amssymb,theorem,dsfont,wasysym}
\usepackage{amssymb}
\usepackage{amsmath}
\usepackage{amstext}
\usepackage{graphicx,epsfig}
\usepackage{epsfig}
\usepackage{verbatim} 
\usepackage{fancyhdr}
\usepackage{fancybox}
\usepackage{color}
\usepackage{ulem,bbold}
\usepackage{enumitem}
\usepackage{subfigure}
\usepackage{bbm}
\usepackage{parskip}
\usepackage[numbers,sort&compress]{natbib}

\definecolor{MyDarkBlue}{rgb}{0.15,0.15,0.45}
\usepackage[linktocpage=true]{hyperref}
\hypersetup{
colorlinks=true,
citecolor=MyDarkBlue,
linkcolor=MyDarkBlue,
urlcolor=MyDarkBlue,
}

\usepackage[numbers,sort&compress]{natbib}
\usepackage{hypernat}

\def\beq{\begin{eqnarray}}
\def\eeq{\end{eqnarray}}

\def\({\left(}
\def\){\right)}

\def\mpl{M_{\rm pl}}

\def\f{\varphi}

\newcommand{\be}{\begin{equation}}
\newcommand{\ee}{\end{equation}}
\newcommand{\la}{\langle}
\newcommand{\ra}{\rangle}
\def\ea{\end{eqnarray}}
\def\ba{\begin{eqnarray}}

\def\beq{\begin{eqnarray}}
\def\eeq{\end{eqnarray}}

\def\({\left(}
\def\){\right)}

\def\mpl{M_{\rm pl}}

\def\p{\partial}
\def\la{\langle}
\def\ra{\rangle}

\def\lsim{\mathrel{\rlap{\lower3pt\hbox{\hskip0pt$\sim$}}
     \raise1pt\hbox{$<$}}}         
\def\gsim{\mathrel{\rlap{\lower4pt\hbox{\hskip1pt$\sim$}}
     \raise1pt\hbox{$>$}}}         

\def\lsim{\mathrel{\rlap{\lower3pt\hbox{\hskip0pt$\sim$}}
     \raise1pt\hbox{$<$}}}         
\def\gsim{\mathrel{\rlap{\lower4pt\hbox{\hskip1pt$\sim$}}
     \raise1pt\hbox{$>$}}}         

\usepackage{amsmath}
\usepackage{amsfonts}
\usepackage{verbatim}
\usepackage{graphicx}
\usepackage{subfigure}
\usepackage{amssymb}
\usepackage[T1]{fontenc}
\usepackage{slashed}

\begin{document}

\renewcommand{\thefootnote}{\fnsymbol{footnote}}

\makeatletter
\@addtoreset{equation}{section}
\makeatother
\renewcommand{\theequation}{\thesection.\arabic{equation}}

\rightline{}
\rightline{}
   \vspace{1.8truecm}


\vspace{10pt}


\begin{center}
{\Large \bf{How Likely are Constituent Quanta to Initiate Inflation?}}
\end{center} 
 \vspace{1truecm}
\thispagestyle{empty} \centerline{{\large  {Lasha  Berezhiani and Mark Trodden}}
}

\vspace{1cm}
\centerline{{\it Center for Particle Cosmology, Department of Physics \& Astronomy, University of Pennsylvania,}}
 \centerline{{\it  209 South 33rd Street, Philadelphia, PA 19104}}
 
 \vspace{1cm}
 
\begin{abstract}

We propose an intuitive framework for studying the problem of initial conditions in slow-roll inflation. In particular, we consider a universe at high, but sub-Planckian energy density and analyze the circumstances under which it is plausible for it to become dominated by inflated patches at late times, without appealing to the idea of self-reproduction. Our approach is based on defining a prior probability distribution for the constituent quanta of the pre-inflationary universe. To test the idea that inflation can begin under very generic circumstances, we make specific -- yet quite general and well grounded -- assumptions on the prior distribution. As a result, we are led to the conclusion that the probability for a given region to ignite inflation at sub-Planckian densities is extremely small. Furthermore, if one chooses to use the enormous volume factor that inflation yields as an appropriate measure, we find that the regions of the universe which started inflating at densities below the self-reproductive threshold nevertheless occupy a negligible physical volume in the present universe as compared to those domains that have never inflated.

\end{abstract}

\newpage
\setcounter{page}{1}

\renewcommand{\thefootnote}{\arabic{footnote}}
\setcounter{footnote}{0}

\linespread{1.1}
\parskip 4pt

\section{Introduction}

According to the standard lore, inflation is capable of producing the entire observable universe from a small homogeneous domain \cite{Guth:1980zm,Linde:1981mu,Albrecht:1982wi}.  If one postulates the inflating universe to emerge from the space-time foam at the Planck scale, as happens in a number of scenarios, then it is unclear how to discuss the likelihood of this process. However, in scenarios in which it is reasonable to consider the non-inflating universe at sub-Planckian energies, we may investigate the circumstances under which it is likely for the universe to become dominated by inflating patches.

This question has prompted a great deal of effort devoted to understanding the question of initial conditions \cite{Hartle:1983ai,Linde:1985ub,Linde:1993xx,Farhi:1986ty,Farhi:1989yr,Vachaspati:1998dy,Kaloper:2002cs,Albrecht:2004ke,Gibbons:2006pa,Schiffrin:2012zf,Ijjas:2013vea,Guth:2013sya,Linde:2014nna,Ijjas:2014nta,Carroll:2014uoa,Mukhanov:2014uwa}. While it is possible that homogeneous domains capable of seeding inflation may be rare in the early universe, it is often argued that the exponential increase in volume that occurs during the inflationary phase may nevertheless result in the physical volume of the late time universe being dominated by patches that underwent inflation. Moreover, if self-reproduction  - {\it eternal inflation} - occurs, then most of the universe would be inflating even today \cite{Linde:1993xx,Creminelli:2008es}, although there remain open questions about this interesting possibility~\cite{Dvali:2013eja,Martinec:2014uva}. 

In this letter, we would like to take a different approach to the problem of quantifying the likelihood of slow-roll inflation without appealing to the tantalizing idea of eternal inflation. Inspired by the recently proposed corpuscular description of the inflationary background \cite{Dvali:2013eja,Dvali:2014ssa}, we propose describing the pre-inflationary universe as a collection of quanta with a certain probability distribution for their wavelengths. As we shall see, this provides considerable constraints on the initial conditions problem of inflation.

The purpose of this work is not to give an account for the origin of the universe; in fact, we are agnostic about this. Instead, we are simply investigating the idea that inflation may begin under very generic circumstances under a set of favorable assumptions about the early universe. In this sense, this work is in the spirit of~\cite{Vachaspati:1998dy}, but approaches the question from the particle physics perspective rather than the general relativistic one. We carry this out by estimating the upper bound on the probability for a given region in the early universe to begin inflating, assuming that certain conditions are met. In particular, our main assumption is that, in a smooth region at sub-Planckian energy density, most of the quanta are expected to have momenta comparable to the energy scale that sets the local energy density.
We then use this bound to find out whether most of the physical volume of the late time universe could have originated from inflation, assuming that inflated domains are rewarded by the exponential volume factor relative to the non-inflated ones. 

We also emphasize that this work is not an attempt to solve the cosmological measure problem or to promote one proposed measure over any others. We are interested in the question of how likely is it for inflation to begin in a high energy (but in the calculable regime) universe, with our assumptions. We could imagine that, when all the relevant physics is understood, and the measure problem solved, it could be that the conditional probability - given that we observe a universe like ours today, what is the likelihood that it arose from a previously inflating patch - is dominated by these inflated patches. We certainly are not in a position to rule this out. But this is, however,  a question we do not know how to answer. We nevertheless think the question we are answering is interesting. Here rather, we merely {\it assume} the globally defined volume measure, since it a particularly simple one and is frequently invoked in favor of inflation. One could, in principle, carry out a similar analysis to the one here for any other choice of measure should one wish.

\section{The Puzzle}
We will phrase the question we wish to address in the context of the simplest model of inflation, focusing on a single massive scalar field, without self-interactions, minimally coupled to gravity, and ignoring all but gravitational and inflaton degrees of freedom. The action is therefore
\beq
S=\int {\rm d} ^4x\sqrt{-g}\left( \frac{\mpl^2}{2} R-\frac{1}{2}g^{\mu\nu}\p_\mu\f\p_\nu\f - \frac{1}{2}m^2 \f^2\right)\, ,
\label{action}
\eeq
where we have defined the reduced Planck mass via Newton's constant $G_N$ by $8\pi G_N = \mpl^{-2}$, and we require  $\mpl\gg M\gg m$, with $M$ denoting the scale of inflation.
 
In order to seed inflation, the scalar field must assume an approximately homogeneous background form $\f(t)$ at least over few Hubble  distances, where the Hubble scale is related to the energy density stored in the inflaton through the Friedmann equation
\beq
H^2=\frac{\rho}{3\mpl^2}\, .
\eeq
The question we would like to ask is whether it is natural to have such a homogeneous background value of the inflaton over the Hubble patch. 

Our approach is to consider the background as a superposition of modes of given wavelengths. In this language, the requirement that the background be homogeneous over a Hubble patch implies that the constituent quanta have super-horizon wavelengths. As a result, the puzzle can be formulated as follows. In order to start inflation at the energy scale $M\equiv \rho^{1/4}\ll\mpl$, the corresponding Hubble patch must be filled with modes of characteristic wavenumber $k <H$. On the other hand, this appears to be puzzling since it is unclear why a given mode would be much softer than the scale $M$ that sets the local energy density. Furthermore, inflation lasts as long as the slow-roll condition is satisfied 
\beq
\frac{m^2}{3H^2}\ll 1\,.
\label{epsilon}
\eeq
Consequently, we must require a hierarchy $m\ll H\ll M$ for accelerated expansion. This means that most regions at such a high energy density may be dominated by sub-horizon modes. In other words, it may be more natural to assume that a given patch of size $H^{-1}$ at energy density $\rho=M^4$ is behaving as a radiation dominated universe rather than an inflating one. Once modes start out with sub-horizon wavelengths, driving decelerated expansion, they will never be able to exit the horizon. 

The following remark is in order here. Sometimes it is argued that even if the energy stored in super-horizon modes is subdominant at the beginning, it may become relevant in time. This could happen since the amplitude of these modes is frozen, while sub-horizon modes decay. However, those quanta that are nearly of the horizon scale will soon reenter the horizon and will themselves decay. As for the modes that are much longer than the curvature scale, their homogeneity will be spoiled quickly, unless the background is fine-tuned to be homogeneous over distances longer than their wavelength. This is an important point and its more detailed analysis will be presented elsewhere.

Notice that if $M$ were of order of $\mpl$ then $H$, $M$ and $\mpl$ would all be of the same order, and that therefore, in this case, homogeneity over the Hubble patch is not as unnatural as it might be at sub-Planckian energies. 

The goal of the remainder of this paper is to quantify the above mentioned unnaturalness.


\section{Estimating Probabilities}

Let us consider the theory \eqref{action} of gravitons and inflatons only. The pre-inflationary state of the universe can be characterized in many different ways, but we would like to think of it in terms of the constituent quanta. Independently of what was happening before inflation, which we assume to start at a scale $M\ll \mpl$, the pre-inflationary processes should have provided us with a region filled with $\f$-quanta. Otherwise, the region in question would not have the potential to inflate.

In order for a given region of size $\ell$ to inflate, the energy density of $\f$-quanta with wavelength $\lambda>\ell$ must be sufficient to source a curvature $H^2\geq\ell^{-2}$. This sets a lower bound on the number of long modes in the region under consideration. In particular, if we denote the average wavenumber as $\la k \ra$, then the total energy density of $N$ such quanta reduces to $N\la k\ra /\ell^3$. As a result, the expression for the curvature sourced by the long modes becomes
\beq
H^2=\frac{N\la k\ra}{3\ell^3 \mpl^2}\,.
\label{curvature}
\eeq
Therefore, the critical number of $\f$-quanta necessary to drive inflation in a region of size $\ell$ is given by
\beq
N\geq N_c\equiv\frac{3\mpl^2\ell}{\la k\ra}\,.
\label{Ncritical}
\eeq

To estimate the probability of inflation we need to be more specific about the initial state. In order to challenge the idea that inflation can begin efficiently under very generic circumstances, we assume that the most likely momentum of a given particle is comparable to the energy scale that sets the local energy density. Moreover, we assume that the probability for a mode to be much softer than this energy scale is significantly smaller than unity. The validity of this assumption depends on the production mechanism of $\f$-quanta. For instance, we could think of the modes in question as the byproduct of the processes with sub-Planckian momentum transfer. In this case, we would expect them to have momenta comparable to the energy scale at which they were produced. Therefore, we propose to consider the wavenumber of a mode to be given by a normalized probability distribution $f(k)$, with the following properties

\begin{enumerate}

\item $f(0)=0$. This assumption is not vital for our argument, as we already mentioned above, all we really need to assume is that the probability for producing ultra-soft modes at high energies is small.

\item $f(k)$ depends on a single energy scale $\la E \ra$ determined by the local energy density via $\rho\equiv \la E \ra ^4$.

\item $f(k)$ has a single maximum around $k\sim \la E \ra \equiv \rho^{1/4}$.

\end{enumerate}

\begin{figure}
\centering
\includegraphics[width=11 cm]{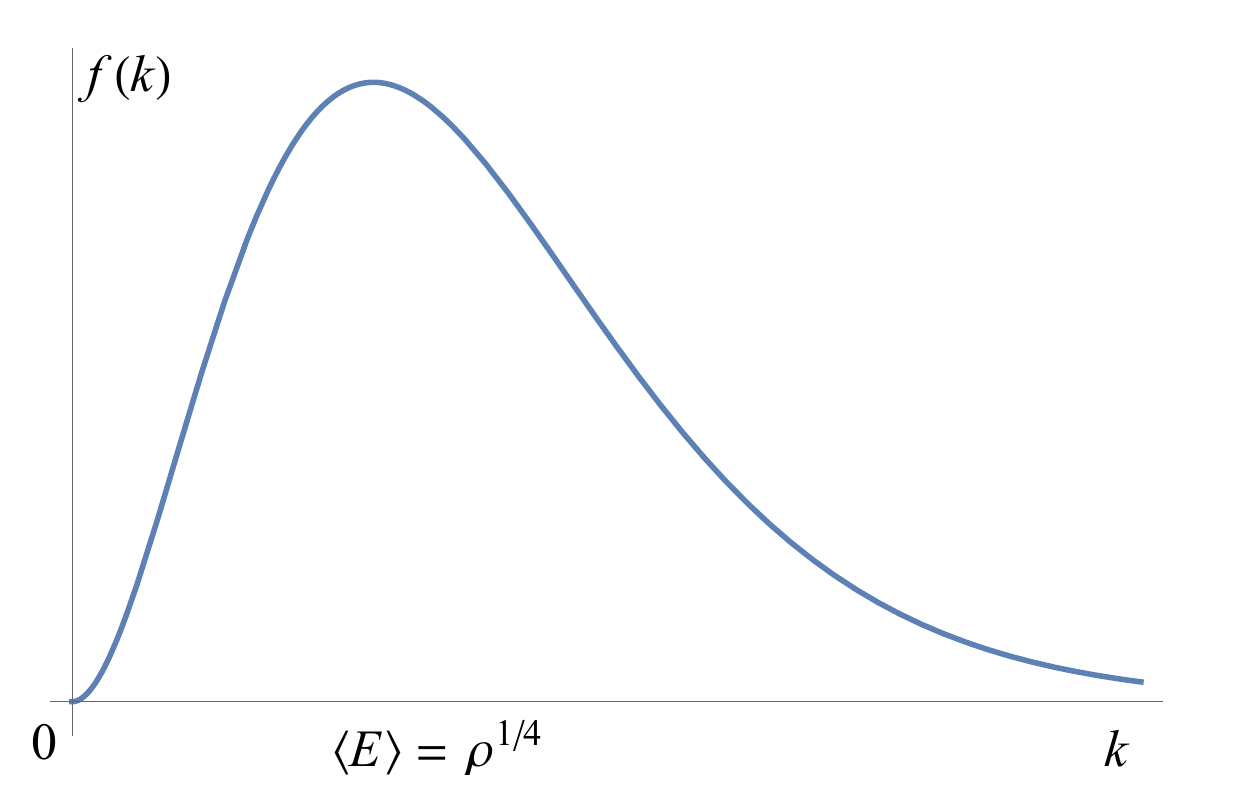}
\caption{The shape of a distribution function which satisfies our requirements.}
\label{fig1}
\end{figure}
A sketch of the distribution function we have in mind is given in Fig. \ref{fig1}. It should be noted that Planck's law for black body radiation is a particular case of the general distribution we consider here.

According to the properties of $f(k)$, the probability for a given particle to have a wavenumber $k\leq \ell^{-1}$ is given by
\beq
p_{\rm soft}=\int _0^{\ell^{-1}} dkf(k)\,.
\eeq
It follows from $\la E \ra \gg H$ that this expression is sensitive to the low $k$ limit  of $f(k)$
\beq
\lim _{k/\la E \ra\to 0} f(k)\propto\frac{k^{\alpha-1}}{\la E \ra ^\alpha}\,,
\label{longf}
\eeq
with $\alpha>1$, resulting in
\beq
p_{\rm soft}\propto \left(\ell\la E \ra\right) ^{-\alpha}\ll 1\,.
\label{Plong}
\eeq

While the constraints on the distribution function we have assumed are particularly convenient for us to work with, it is likely that they could be relaxed without qualitatively changing our conclusion, as long as we ensure that $p_{\rm soft}\ll1$.

We now establish the fraction of quanta that we require to be soft in order to launch inflation. As we have already mentioned above, the total energy in modes with $k<\ell^{-1}$ must be much larger than that in modes with $k>\ell^{-1}$. We proceed by noticing that the mean momentum for the super-$\ell$ and sub-$\ell$ modes are, respectively, given by
\beq
\langle k \rangle _{k\ell>1} =\frac{\int_0^{1/\ell} dkkf(k)}{\int_0^{1/\ell} dkf(k)}\sim \ell^{-1}\,, \qquad \langle k \rangle _{k\ell<1} =\frac{\int_{1/\ell}^{\la E \ra} dkkf(k)}{\int_{1/\ell}^{\la E \ra} dkf(k)}\sim \la E \ra \,.
\label{avg}
\eeq
Requiring $N_{k\ell>1}/\ell\gg N_{k\ell<1} \la E \ra$ and that the curvature radius be smaller than $\ell$, we obtain that the number of short modes must satisfy
\beq
N_{k\ell<1}\ll \frac{N_{k\ell>1}}{\sqrt{\ell \mpl}}\,.
\label{fraction}
\eeq

Again, the region in question may inflate if it contains a sufficient number of quanta and if a significant fraction \eqref{fraction} of these quanta are longer than the size of the region. However, the bound \eqref{fraction} is so tight that we can simply demand that all the modes be long, making a negligible error in the final answer while simplifying the discussion.

As a result, we proceed by computing the probability of inflation as
\beq
\mathcal{P}_{inf}(H)=\mathcal{P}_{N} \cdot p_{\rm soft}^{N}\, .
\label{probability}
\eeq
Here $N$ denotes the number of modes of wavelength longer than $\ell$ needed in order to drive inflation in a region of size $\ell$ with the curvature scale $H$. In \eqref{probability}, $\mathcal{P}_{N}$ stands for the probability that the region in question contains a number $N$ of $\f$-quanta, and $p_{\rm soft}^{N}$ is the probability that all of these quanta have wavelengths larger than the size of the region. Notice that $p_{\rm soft}$ is determined by the local characteristics of the region at hand and is independent of the global properties of the universe. The first factor of \eqref{probability}, on the other hand, is determined by the distribution function of the entire universe \footnote{In order to explain what we mean under this,  let us imagine a large room filled with air. Then the probability of finding a certain number of air molecules in a small volume inside the room depends, for instance, on a total number of molecules in the room.}. In the appendix we will consider the example of a special initial state for which we will be able to quantify $\mathcal{P}_{N}$.

In order to establish an upper bound on $\mathcal{P}_{inf}$, let us overestimate it in favor of inflation occurring. For this, notice that $\mathcal{P}_{N}<1$ which leads to
\beq
\mathcal{P}_{inf}(H)< p_{\rm soft}^{N}\,.
\label{Pbound}
\eeq
Taking into account \eqref{avg}, the expressions for the curvature \eqref{curvature} and the critical number \eqref{Ncritical} reduce to
\beq
H^2=\frac{N}{3\ell^4\mpl^2}\,, \qquad N\geq N_c\equiv 3\mpl^2\ell^2\,,
\label{HN}
\eeq
and using \eqref{Plong}, \eqref{Pbound} this can be recast as
\beq
\mathcal{P}_{inf}(N\geq N_c)<(\ell\la E \ra)^{-\alpha N}\,.
\eeq
This expression may be further simplified. To do this, let us remind ourselves that $\la E \ra$ is the expected energy of the mode, which is determined in terms of the local energy density. If we have $N$ quanta in a region of size $\ell$, then $\la E \ra$ can be determined as follows
\beq
\frac{N \la E \ra}{\ell^3}=\la E \ra ^4\,, \qquad \Rightarrow \qquad \la E \ra=\frac{N^{1/3}}{\ell}\,.
\eeq
As a result we obtain the following bound on the probability of inflation
\beq
\mathcal{P}_{inf}(N\geq N_c)< N^{-\alpha N/3}\,.
\label{finprob}
\eeq
Notice that, since $N\gg 1$, $\mathcal{P}_{inf}$ is exponentially small. Therefore, at the onset of inflation the homogeneous regions capable of seeding accelerated expansion are extremely rare.

However, the fact that the probability for a given region to inflate is exponentially small is not necessarily problematic. Considering that the regions that do inflate are stretched over the exponentially larger physical volume, the universe could be still dominated by inflated regions after reheating.

To be more specific, the physical volume of a region at the end of inflation $V_f$ is related to the initial volume $V_i$ by
\beq
V_f=V_i e^{3 \mathcal{N}}\,.
\eeq 
Here, $\mathcal{N}$ stands for the number of e-folds from the beginning of inflation to the end. Therefore, the necessary condition for the universe to become dominated by inflated patches at late times is given by
\beq
\mathcal{P}_{inf} e^{3 \mathcal{N}}>1\,.
\label{condition}
\eeq
Assuming $\alpha/3\sim 1$ and ${\rm ln}N\gsim 1$ in \eqref{finprob}, it is easy to see that for \eqref{condition} we need
\beq
\mathcal{N}>N\,.
\label{efolds}
\eeq
For the model under consideration, this inequality implies the following bound on the energy density
\beq
\rho> \mpl^3 m\,.
\eeq
Interestingly, this bound coincides with the scale of eternal inflation for the $m^2\f^2$ model. Therefore, domains that started inflating below the self-reproduction threshold occupy a negligible volume compared to those domains that never inflated.

We would like to point out that according to \cite{Dvali:2013eja}, the maximum number of e-folds, after which the semi-classical description is expected to break down, satisfies
\beq
\mathcal{N}_{\rm max}<N^{2/3}\,.
\eeq
Comparing this to our result~\eqref{efolds} we see that requiring that we remain in the semi-classical regime conflicts with the requirement that the universe be dominated by inflated patches at late times.

We conclude this section by pointing out a few caveats. The central expression of our work is \eqref{Pbound}. Therefore, the likelihood of inflation is determined by the magnitude of $p_{\rm soft}$. So far we have ignored condensation effects, since we have been demanding the modes have super-horizon wavelengths. Instead, we might wonder whether modes need to be just soft enough that they overlap significantly. For this to be true, the characteristic wavelength of quanta must be much longer than the inter-particle separation. If $k$ denotes the wavenumber of quanta and $\la E \ra ^{-1}$ is the inter-particle separation given by \eqref{avg}, this implies
\beq
k\ll\la E\ra\,.
\label{condensate}
\eeq
In thermodynamic terms this means that the temperature $T$ must be significantly lower than the critical temperature. Moreover, the quanta in the condensate will carry energy comparable to $m$, which is the smallest energy scale at hand. Therefore, in order for the energy stored in the condensate to dominate over the energy of the normal component, it is necessary that, as given in~\eqref{condensate}, $k\ll\la E\ra$ and not merely that $k<\la E\ra$.

However, in order for condensation to take place,  it is necessary that the domain in question be in equilibrium, which could be achieved if the rate of interactions is greater than the rate of cosmological evolution. Unsurprisingly, this is quite difficult to achieve, since gravitational interactions are weak at sub-Planckian energies. In fact, it follows from a simple estimate that the universe is expected to be out of equilibrium on average. However, a region filled with quanta of the characteristic wavenumber $k$ could reach equilibrium if the following condition is satisfied
\beq
\frac{k}{\la E \ra}\ll \frac{\la E\ra^2}{\mpl^2}\,.
\eeq
This is a stronger requirement than just requiring a significant overlap and, as a result, we are once again led to $p_{\rm soft}\ll 1$. Therefore, a good estimate for the bound on the probability of inflation is
\beq
\mathcal{P}_{inf}<e^{-N}\,,
\eeq
which is precisely what we used to argue \eqref{efolds}.


\section{Conclusions}
The initial conditions question of inflation is an important component of understanding the extent to which the paradigm solves the problems of the standard big bang cosmology. It is possible that eternal inflation might ameliorate any such problems, allowing for the probability that a given FRW-like patch of the late universe arose from inflation to be arbitrarily close to unity. However, eternal inflation may not take place in a given model, and there remain open questions about the general idea, with a number of authors raising arguments that complicate the standard picture. 

In this paper we have approached the initial conditions problem in a new way, inspired by the recently proposed corpuscular description of the inflationary background \cite{Dvali:2013eja,Dvali:2014ssa}. We have proposed describing the pre-inflationary universe as a collection of quanta, and have rephrased the usual issues as questions about the probability distribution of the wavelengths of these quanta. We have shown that seemingly reasonable assumptions about this probability distribution result in the exponentially small probability for the beginning of inflation. Even after taking into account the exponential volume factor that inflation yields, we find that those regions that started inflating below the self-reproduction threshold occupy a negligible volume in the present universe as compared to the regions that have never inflated. Therefore, the problem of initial conditions cannot be solved by modifying the potential in such a way that the Planckian densities are reached without ever entering the self-reproduction regime. We would like to stress that although we have focussed on $m^2\f^2$ model, we believe that the modification of the potential should not change our conclusion about slow-roll inflation, unless significant fine-tuning is involved. False-vacuum inflation, however, may require a separate study.

We therefore conclude that inflation does not begin efficiently under the assumptions we have laid out, and that special arrangements of the initial conditions are required. 

\vspace{0.5 in}

{\bf Acknowledgements:} We are especially grateful to Sean Carroll, Gia Dvali and Justin Khoury for illuminating discussions and comments. We would also like to thank Gregory Gabadadze, Gary Gibbons, Matthew Kleban, Paul Steinhardt and  Tanmay Vachaspati for useful discussions. L.B. is supported by funds provided by the University of Pennsylvania. The work of M.T. is supported by US Department of Energy grant DE-FG02-95ER40893, and by NASA ATP grant NNX11AI95G. 

\section*{Appendix: Modeling the Initial State of the Entire Universe}
\renewcommand{\theequation}{A-\Roman{equation}}
\setcounter{equation}{0}

In this appendix, we give a specific special example of how we might model the initial state of the entire universe. The state we consider here shares some similarities with the one used in section 3, but there are significant differences as well. 

Let us consider a gas of $\phi$-particles with the following properties at energy scale $M$:

i) The wavenumber of a particle is given by a normalized probability distribution $f(k)$, which peaks around $k=M$ and vanishes for $k=0$.

ii) The number density of particles is uniformly distributed throughout the infinite volume of the universe. In other words, we assume translation invariance is unbroken.

Notice that here, the expectation value of a wavenumber of a given mode is assumed to be of order $M$, regardless of the local environment. This is the point that distinguishes this state from the one considered in section 3. There, the expected momentum depended on how densely packed a region was.

On average, a universe which satisfies these conditions will be filled with relativistic quanta and will have a characteristic Hubble scale
\beq
H=\frac{M^2}{\mpl}\,.
\label{aH}
\eeq
In order to regularize the infinite volume we consider a box of size $L\gg H^{-1}$, containing a fixed number of particles
\beq
N_L=M^3L^3\, ,
\eeq
keeping in mind that we eventually intend to take the $L\rightarrow\infty$ limit.

Now, we can use the probability distribution specified above to compute the probability of inflation at any scale. For definiteness, let us compute the probability for a given region to start inflating at energy density $\rho=M^4$ and thus with the Hubble scale \eqref{aH}. In other words, the question we would like to ask is what is the probability for a given box of volume $H^{-3}$, located inside a box of size $L$, to be sufficiently homogeneous to seed  inflation? The probability for a given particle to be located inside a box of size $H^{-1}$ is
\beq
p_1=\frac{1}{H^3L^3}\, ,
\eeq
and the probability of having a number $N_H$ of $\phi$-particles inside a given Hubble volume is given by the binomial distribution
\beq
\mathcal{P}_{N_H}=\frac{N_L!}{N_H!(N_L-N_H)!}p_1^{N_H}(1-p_1)^{N_L-N_H}\, .
\eeq
Notice that in the $L\to \infty$ limit, $N\to \infty$ and $N_L p_1=M^3/H^3$ is fixed. As a result, in the limit of the infinite universe, the binomial distribution converges to the Poisson distribution
\beq
\mathcal{P}_{N_H}=\frac{(M/H)^{3N_H}}{N_H!}e^{-(M/H)^3}\,.
\label{poisson}
\eeq
Further, since we are working in the $M\ll\mpl$ limit, we have $M/H\gg 1$ and the above probability distribution can then be approximated by the normal distribution 
\beq
\mathcal{P}_{N_H}=\frac{1}{\sigma\sqrt{2\pi}}e^{-\frac{(N_H-\sigma)^2}{2\sigma^2}}\,,
\eeq
with $\sigma^2\equiv(M/H)^3$.

In order for the Hubble patch to begin inflating, it is necessary for it to be sufficiently homogeneous. In particular, most of the energy inside the box should come from particles of wavenumber $k\leq H$; moreover, the energy density of these long wavelength particles should be at least $\rho=M^4$ in order to yield a curvature radius at most $H^{-1}$.

According to property (i), the probability for a given particle to have a wavenumber smaller than $H$ is given by
\beq
p_{\rm soft}=\int _0^H dkf(k)\,.
\eeq
As before, it follows from $M\gg H$ that the probability for a given particle to have a wavenumber $k\leq H$ is sensitive to the low $k$ limit  of $f(k)$
\beq
\lim _{k/M\to 0} f(k)\propto\frac{k^{\alpha-1}}{M^\alpha}\,,
\label{longf}
\eeq
with $\alpha>1$. Notice that we assume vanishing probability for having strictly $k=0$ modes. 

Now, let us establish the critical number of particles sufficient to launch inflation. As we have already mentioned above, the energy density in the modes with $k<H$ must be at least $M^4$. To calculate this critical number we must find the mean momentum for the super-horizon modes
\beq
\langle k \rangle _H =\frac{\int_0^H dkkf(k)}{\int_0^H dkf(k)}\sim H\,.
\eeq
Hence, for the critical number we obtain $N_{c}H^4=M^4$, which gives
\beq
N_{c}=\frac{M^4}{H^4}\,.
\eeq

In analogy with section 3, let us demand that the Hubble patch under consideration is absolutely homogeneous, i.e. that all the modes inside are longer than the curvature radius. Then the probability for a given region to be homogeneous enough to seed inflation is given by 
\beq
\mathcal{P}_{inf}=\mathcal{P}_{N_H} p_{\rm soft}^{N_H}\,.
\eeq
Furthermore, in order for inflation even to begin we need $N_H>N_{c}$, as we have already mentioned above. It is easy to see that this probability is extremely suppressed, just as in section 3. In fact, all the conclusions of section 3 hold here as well.

\end{document}